\documentclass{article}
\usepackage{spconf,amsmath,graphicx,hyperref}
\usepackage{booktabs} 
\usepackage{amssymb}

\usepackage{xcolor}
\definecolor{darkred}{rgb}{0.6,0.0,0.0}
\definecolor{darkgreen}{rgb}{0.0,0.6,0.0}

\title{Beyond Video-to-SFX: Video to Audio Synthesis with Environmentally Aware Speech}
\name{Xinlei Niu$^{1}$\thanks{Work done during the internship at Dolby.}, Jianbo Ma$^2$, Dylan Harper-Harris$^2$, Xiangyu Zhang$^{3}$, Charles Patrick Martin$^1$, Jing Zhang$^1$}
\address{$^1$ Australian National University; $^2$ Dolby Laboratories; $^3$ The University of New South Wales}

\begin{document}
\ninept
\maketitle
\begin{abstract}
The generation of realistic, context-aware audio is important in real-world applications such as video game development. While existing video-to-audio (V2A) methods mainly focus on Foley sound generation, they struggle to produce intelligible speech. Meanwhile, current environmental speech synthesis approaches remain text-driven and fail to temporally align with dynamic video content. In this paper, we propose Beyond Video-to-SFX (BVS), a method to generate synchronized audio with environmentally aware intelligible speech for given videos. We introduce a two-stage modeling method: (1) stage one is a video-guided audio semantic (V2AS) model to predict unified audio semantic tokens conditioned on phonetic cues; (2) stage two is a video-conditioned semantic-to-acoustic (VS2A) model that refines semantic tokens into detailed acoustic tokens. Experiments demonstrate the effectiveness of BVS in scenarios such as video-to-context-aware speech synthesis and immersive audio background conversion, with ablation studies further validating our design. Our demonstration is available at~\href{https://xinleiniu.github.io/BVS-demo/}{BVS-Demo}.
\end{abstract}
\vspace{-5pt}
\section{Introduction}
\label{sec:intro}
\vspace{-4pt}
The generation of realistic, high-fidelity audio, including sound effects and speech, has become increasingly important for real-world applications such as film production, virtual reality (VR), and game development. To this end, useful generative models must generate audio that is coherent, expressive, and contextually aligned with the content of the given condition. Within this domain, video-to-audio (V2A) focuses on generating audio that is temporally and semantically aligned with visual cues in videos. In many scenarios, the audio track includes not only visually grounded sounds but also speech, and the speaker may not be directly observable. For example, a video clip of someone washing dishes might include background conversations that are not visually depicted. Motivated by this challenge, our work aims to develop a method capable of synthesizing audio that incorporates environmentally aware speech. The goal is to generate audio that is temporally aligned with the video while ensuring semantic consistency and incorporating context-aware spoken content.

Previous V2A studies~\cite{luo2023diff,wang2024frieren,xue2024auffusion,lee2024video,chen2024fastsag,viertola2025temporally} focus on Foley sound generation, aiming to synthesize ambient sounds that are temporally and semantically aligned with the video. More recent approaches~\cite{liu2024tell,cheng2025mmaudio} extend this paradigm by incorporating text prompts alongside video input. The inclusion of textual context enables models to capture more complex scenarios, such as the sound scene of `a man speaking while he is playing tennis'. However, these methods remain limited to sound event-level alignment driven by visual and/or textual cues and lack the capability to generate coherent and intelligible speech.
Consequently, they often fail to generate natural spoken content, which restricts their real-world application where context-aware spoken content is essential.

A closely related task is video-to-speech (V2S)~\cite{choi2025v2sflow,liang2025lightl2s}, which focuses on generating speech from videos with talking heads. However, V2S methods primarily address clean speech synthesis and do not handle sound effect generation, making them complementary but distinct from V2A approaches. A concurrent work by~\cite{tian2025dualdub} demonstrates V2S generation with integrated sound effects; however, their approach is primarily designed for videos with talking heads.
In parallel, environmental context-aware speech synthesis~\cite{lee2024voiceldm,jung2025voicedit} focuses on generating speech that incorporates background noise consistent with the given text prompts. These methods typically leverage a complex training strategy to learn the speech information. They first pretrain on synthetic noisy speech datasets and then fine-tune on real-world datasets transcribed by automatic speech recognition (ASR) models to improve speech modeling. Other related approaches~\cite{glazer2025umbratts,he2025multi} explore mixing clean speech with ambient sounds to create contextually consistent audio environments. While these techniques are effective in achieving semantic consistency between speech and background context, they generally lack fine-grained temporal control over sound events, a capability that is critical for synchronizing sound effects with dynamic video content.

\begin{figure}[t]
\vspace{-0.03in}
\centerline{\includegraphics[width=0.8\linewidth]{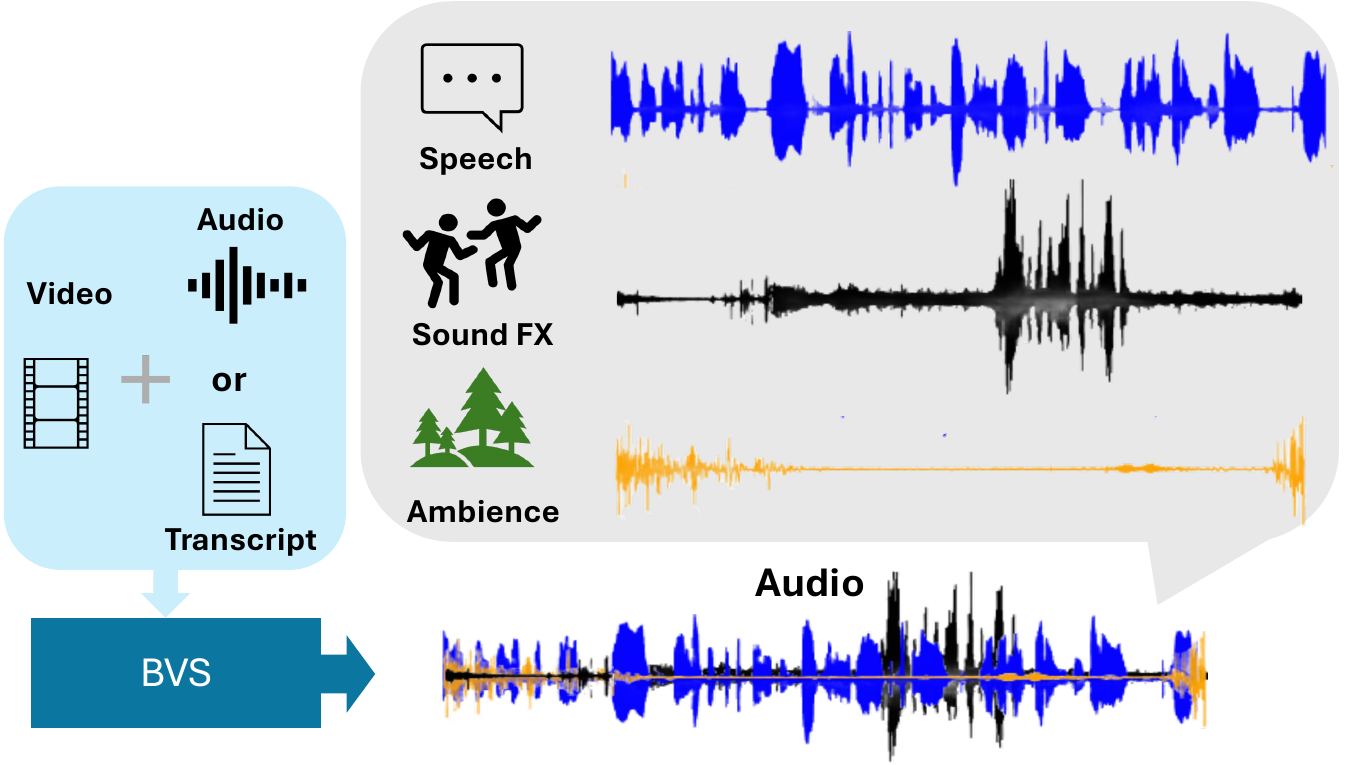}}
\vspace{-0.2in}
\caption{The proposed BVS framework.}
\vspace{-0.25in}
\label{fig:surface}
\end{figure}

\begin{figure*}[t]
\vspace{-0.15in}
\centerline{\includegraphics[width=0.96\textwidth]{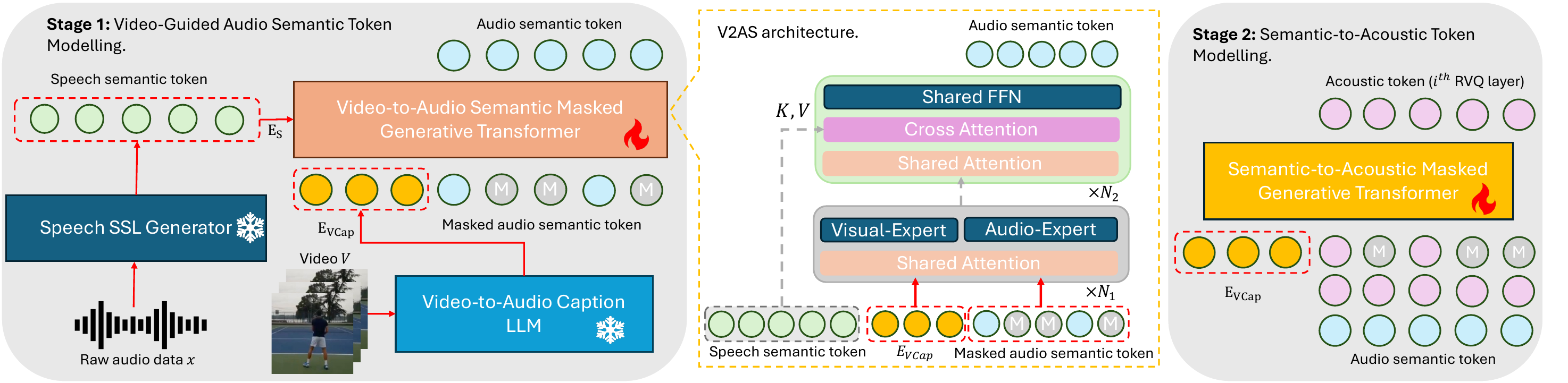}}
\vspace{-0.15in}
\caption{Overview of the BVS training pipeline. In two stages, BVS generates unified audio semantic tokens from video and phonetic cues, then refines them into acoustic tokens. Snowflakes mark frozen pretrained models, and fire icons represent trainable components.}
\label{fig:overview}
\vspace{-0.2in}
\end{figure*}

We identify two major limitations in previous works. Firstly, a key challenge is the scarcity of speech datasets that simultaneously provide context-rich environmental sounds, paired video streams, and ground-truth transcripts. Previous methods~\cite{lee2024voiceldm,jung2025voicedit} first train their models with synthesized noisy speech by mixing clean speech data for TTS training with environmental sound data, and then fine-tune using real-world unlabeled audio data containing speech with transcripts obtained from ASR models.
However, ASR transcripts often suffer from inaccuracies and unstable segmentation in noisy speech, which introduces incorrect labels during speech learning. Meanwhile, synthetic noisy speech lacks natural video grounding since corresponding video streams cannot be created, limiting multimodal consistency. Secondly, current V2A models often fail to generate intelligible and natural speech, and often lack control over the speech content. As a result, additional post-production is often needed to remove unintelligible vocals and mix clean speech with ambient sounds for contextual alignment, which reduces their practicality in real-world applications. 

In this paper, we propose Beyond Video-to-SFX (BVS), a method for synthesizing video-synchronized audio that incorporates environmental-aware speech. As shown in Fig.~\ref{fig:surface}, BVS generates soundtracks that are temporally and semantically aligned with visual content while integrating context-aware intelligible speech, guided by provided transcripts or audios containing spoken content.
We make the following contributions:
(1) We propose BVS, an improved V2A framework enabling intelligible speech synthesis, addressing a core limitation of previous approaches.
(2) Our method is trained directly on raw audio data, eliminating the need for complex preprocessing steps such as synthetic noisy datasets or filtering ASR-generated transcripts.
(3) We demonstrate that BVS is a flexible framework for various downstream tasks, including audio background conversion and video to context-aware speech synthesis.

\section{Method}\label{sec:method}

Generating real-world audio that matches a visual scene is challenging. Audio tracks such as speech, sound effects, and ambient sounds are each influenced by distinct factors. For example, speech often occurs with irregular onset and offset times. It is primarily human-related, frequently off-screen, and relies more on phonetic cues than on visual context. In contrast, sound effects and ambient sounds are closely linked to visual objects and scene context, exhibiting highly variable temporal alignment with the video. These fundamental differences make it difficult for a single model to jointly generate both intelligible speech and natural ambient sounds directly from phonetic and visual cues. Moreover, to create a convincing audio experience for a given video, all audio tracks must remain coherent and share consistent environmental characteristics.

To address this, we introduce an audio semantic token space as a unified representation of speech and ambient sounds inspired from two-stage TTS pipelines~\cite{wang2024maskgct,wang2025metis}. The semantic tokens capture abstract audio information, reducing the gap between multi-modal conditions and acoustic outputs. As shown in Fig.~\ref{fig:overview}, our BVS pipeline operates in two stages: (1) fuse information across speech, visual, and contextual modalities into audio semantic tokens, and (2) refine these audio semantic tokens into acoustic outputs. This separation reduces the burden of a single-stage model that needs to learn both multimodal fusion and acoustic generation simultaneously.

\begin{figure}[t]
\centerline{\includegraphics[width=\linewidth]{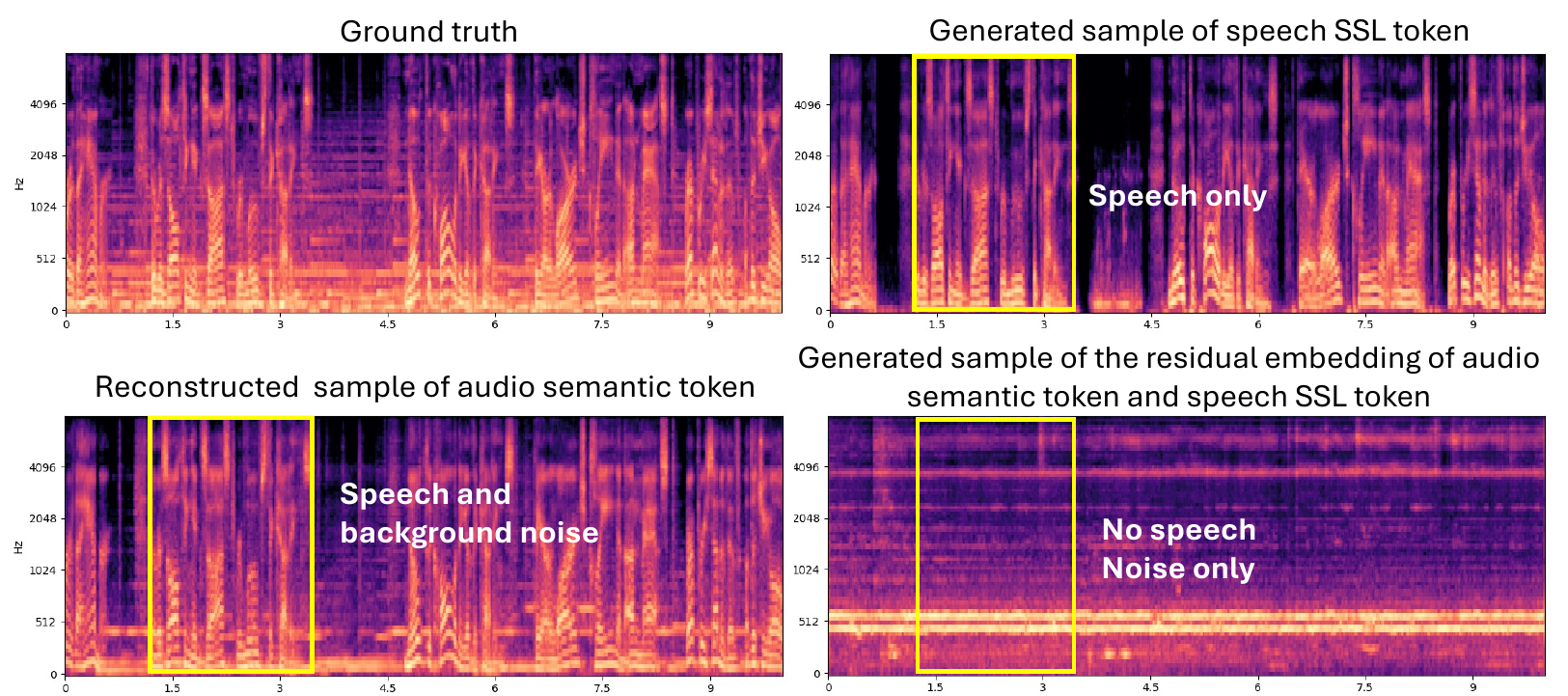}}
\vspace{-0.1in}
\caption{The generated sample of residual SSL highlights background noise, indicating sound effect is independent to speech information.}
\vspace{-0.2in}
\label{fig:residual} 
\end{figure}

\vspace{-3pt}
\subsection{Video-Guided Audio Semantic Token Modeling}
\label{sec:v2a-semantic-generative-modeling}
We design a video-to-audio semantic (V2AS) masked generative transformer, which generates audio semantic tokens by jointly integrating the speech and sound effects from video and phonetic cues.  As mentioned before, speech, sound effects, and ambient sounds are each related to different objects. We split the audio semantic token generation into two components: (1) clean speech semantic token generation in the form of speech SSL token; (2) audio semantic fusion according to speech and video cues.

\noindent\textbf{Noisy Audio to Speech Semantic Token:} 
We aim to mitigate inaccurate transcription in previous methods~\cite{lee2024voiceldm,jung2025voicedit} and directly train models on raw noisy audios containing speech. To guide the model in identifying speech regions within raw audio,
we leverage a pretrained speech SSL token generator~\cite{wang2025metis} trained for the speech enhancement (SE) task to extract speech information from noisy inputs. Rather than employing the full model to directly separate clean speech waveform, we use its first-stage generative transformer to predict speech self-supervised learning (SSL) token sequences in the quantized W2V-BERT feature space~\cite{chung2021w2v}, obtained via a VQ-VAE~\cite{wang2024maskgct}. Formally, given a raw audio input $x$, the pretrained SE masked generative transformer captures speech SSL tokens  $S_s$ from $p_\phi(S_s|x)$, where $\phi$ denotes the parameter of the pretrained SE model.
This design emphasizes capturing of speech information, rather than reconstructing the clean speech waveform, which offers two key advantages: (1) it disentangles speech information more effectively under noisy conditions of the raw audio, and (2) it reduces downstream decoding errors by using compact speech SSL tokens.

\noindent\textbf{Audio Semantic Fusion:} 
While speech semantic tokens allow the model to be aware of speech regions directly from raw audio, the remaining challenge is to fuse the speech information with video context while generating synchronized audios from visual cues. To address this, we make use of the quantized W2V-BERT feature space as the unified audio semantic token space, which is motivated by our empirical findings: Although trained primarily on speech, SSL models also encode non-speech acoustic cues. This is also pointed out in~\cite{chang2025usad}. We leveraged a pretrained semantic to acoustic generative model to reconstruct the waveform for a given SSL token sequence including both speech and background noise. We empirically observed that the reconstructed audio contains both intelligible speech and matching background noise. Fig.~\ref{fig:residual} further illustrates this empirical finding. We also provide the audio samples on our ~\href{https://xinleiniu.github.io/BVS-demo/}{Demo-page}. Reconstructions of the W2V-BERT SSL tokens preserve both speech and non-speech information. Our empirical analysis further shows that the residual difference between speech SSL embeddings and full audio semantic embeddings potentially corresponds to non-speech components such as background noise or ambient sounds. This suggests that speech and non-speech information can be disentangled and recombined in the SSL embedding space.
Motivated by this observation, the V2AS model predicts audio semantic token sequences in the same quantized W2V-BERT feature space that fuses non-speech acoustic information with speech information into audio semantic tokens using complementary cues from speech SSL tokens and videos. This process is shown in the right-hand side of stage 1 in Fig.~\ref{fig:overview}.
By predicting semantic tokens rather than raw acoustic outputs, the model focuses on learning cross-modal relationships between speech, ambient sounds, and visual context.

\noindent\textbf{Training Objective:} 
We use the mask-and-predict learning strategy from~\cite{chang2022maskgit} and mask the ground truth audio semantic token sequence $S_a$ at each time step $t$ with a corresponding binary mask $M_t$. Items in $S_{a}$ are replaced with a special [MASK] token if $m_{t}^i = 1$, otherwise $S_a^i$ remain unchanged. The resulting masked token sequence is denoted as $S_{a}^{M}$. Each $m_t^i$ is i.i.d to a Bernoulli distribution with parameter $\gamma(t)\in(0,1]$. Thus, the V2AS model is formulated as $p_{\theta_{\text{V2AS}}}(S_a|S^M_a,V,S_s)$, where $V$ represents video conditions. We train the model using cross-entropy loss between predicted audio semantic tokens $\hat{S}_a$ and ground truth $S_a$ as 
\begin{equation}
\mathcal{L}_{\text{V2AS}} = - \sum_{s \in S_a^M} \mathbb{I}(s = [\text{MASK}])\log[P_{\theta} (\hat{s} = s| S_a,S_s, V)] 
\end{equation}
\noindent\textbf{Architecture:} 
To further enhance the model's video understanding ability, we incorporate visual representations enriched by audio captions generated with a V2Cap large language model (LLM)~\cite{liu2024tell}, which we denote as $E_{\text{VCap}}$. The detailed V2AS architecture is shown in Fig.~\ref{fig:overview} (in the yellow dash box), which extends a V2A transformer framework~\cite{liu2024tell} by injecting speech SSL embeddings into extra cross-attention blocks as $K$ and $V$, where $N_1 = 12$ and $N_2 = 14$. This integration improves speech localization and intelligibility, and captures non-speech details guided by video context. During training, we concatenate the masked audio semantic embeddings $E_a^M$ and $E_{\text{VCap}}$ along the temporal axis. During inference, the V2AS model generates audio semantic token sequences conditioned on the video prompt and speech SSL tokens. The speech SSL token sequences can be obtained by a unified speech generative transformer such as Metis stage 1~\cite{wang2025metis}, which takes transcripts, audio, or other inputs.

\subsection{Semantic-to-Acoustic Token Modeling}

We then train a video-guided semantic-to-acoustic (VS2A) model to obtain acoustic token sequences conditioned on audio semantic token sequences and $E_{\text{VCap}}$. The VS2A model refines generation by mapping abstract audio semantic representations into acoustic tokens while leveraging video context to enhance the alignment. Following~\cite{wang2024maskgct}, the VS2A model is a masked generative transformer that produces a multi-layer acoustic $A^{1:K}$, where $K$ denotes the number of codebooks, using iterative parallel decoding within each layer. For simplicity, we adopt separate VS2A models: one dedicated to predicting the first-layer acoustic codec and another for the subsequent layers. We denote the $i$-th acoustic token layer as $A^i$, and formulate the model as $p_{\theta_{\text{VS2A}}}(A^i|V,A^{1:i-1})$. During training, we simply sum the embeddings of the audio semantic tokens and the embeddings of the acoustic codec tokens from layer 1 to $i$, which is denoted as $E_{C_i}^M$ . Then we concatenate $E_{\text{VCap}}$ and $E_{C_i}^M$ as input of the model. We use cross-entropy loss between predicted acoustic tokens and ground truth acoustic tokens as
\begin{equation}
\vspace{-3pt}
    \mathcal{L}_{\text{VS2A}} = - \sum_{a^i \in A^i} \mathbb{I}(a^i = [\text{MASK}])\log[P_\theta (\hat{a}^i = a^i| S_a, V)]
    \vspace{-3pt}
\end{equation}
During inference, the VS2A model generates acoustic tokens progressively from coarse to fine across layers, employing iterative parallel decoding within each layer.
\section{Experiment}
\noindent\textbf{Dataset:} 
We train our model on the filtered English subset of AudioSet-Speech~\cite{audioset,lee2024voiceldm} (AS-Speech), which contains approximately 580k audio–video pairs. Each sample has a 10-second duration and contains both speech and various environmental sound events. For evaluation, we adopt AC-filtered~\footnote{\href{https://github.com/glory20h/voiceldm-data}{https://github.com/glory20h/voiceldm-data}}, which includes audio sourced from AudioCap~\cite{kim2019audiocaps}, captions, and transcripts. As AudioCaps is derived from AudioSet, we retrieve the corresponding video streams via the YouTube IDs, resulting in a curated set of 152 samples, which we refer to as AS-filtered dataset.

\noindent\textbf{Implementation Details:}
We train V2AS and VS2A models on three NVIDIA A100 (80GB) GPUs for four and three days, respectively, using a batch size of 32, a learning rate of $2\times10^{-4}$, 20k and 32k warm-up steps, for approximately 80 and 34 epochs, respectively. We optimize these models with the AdamW~\cite{loshchilov2017decoupled} optimizer. We adopt VATT~\cite{liu2024tell} transformer with extra cross-attention blocks for our V2AS model, and we adopt the S2A model architecture in~\cite{wang2024maskgct} for the VS2A model. For the video modality, we employ the EVA-CLIP image encoder~\cite{sun2023eva} to extract mean-pooled frame features sampled at 5 fps, yielding a visual sequence of size $50 \times 768$ for a 10-second video. Following~\cite{liu2024tell}, the visual sequence passes through a pretrained V2Cap LLaMa and we extract its hidden state as input features. We add two learnable modality-specific embeddings to the concatenated inputs in the V2AS model. We use Metis SE stage 1 model to obtain speech SSL tokens~\cite{wang2025metis}. Audio SSL features are extracted from W2V-BERT 2.0~\cite{chung2021w2v}, which are then quantized into discrete tokens at 50 tokens rate per second by the pretrained VQ-VAE in~\cite{wang2024maskgct}. Acoustic tokens are obtained from 16kHz Encodec~\cite{defossez2022high} with four codebooks at a 50 Hz token rate. During inference, we use 16 steps for the V2AS model and [20, 10, 1, 1] steps for the VS2A model. The classifier-free guidance (CFG) scale is set to 5.0 for V2AS and 2.5 for VS2A.

\noindent\textbf{Evaluation Metrics:}
We evaluate our method along three key dimensions: speech intelligibility, temporal alignment, and semantic perceptual similarity. For speech intelligibility, we report both the word error rate (WER) and $\Delta$WER. WER is computed between the ground truth transcripts and those obtained from transcribing our generated audio using the whisper-large-v3. Since ASR models are trained on transcribing clean speech, it may produce inaccurate transcripts on noisy inputs. We additionally report $\Delta\textnormal{WER}$, which measures the absolute difference in WER between the ground truth and generated results, as $\Delta \text{WER} = \frac{1}{N}\sum^N_i|\text{WER}(\text{gt}_i,t_i)-\text{WER}(\text{pred}_i,t_i)|$, where $t_i$ represents the transcripts for audio sample $i$.
To assess temporal alignment between video and generated audio, we follow~\cite{cheng2025mmaudio} and report the DeSync~\cite{iashin2024synchformer}. For semantic and perceptual similarity, we follow~\cite{manor2024zero} and calculate the FAD and LPAPS scores, which are computed using the clap-laion-audio model in music-speech-audioset.

\subsection{Video to Audio with Context-Aware Speech}~\label{sec:tts}
\begin{table}[t]
    \vspace{-0.15in}
    \centering
    {\fontsize{8pt}{9pt}\selectfont
    \renewcommand{\tabcolsep}{0.8mm}
    \begin{tabular}{c|c|cc|cccc}
         \toprule
         Method & Dataset& WER $\downarrow$ & $\Delta$WER $\downarrow$ &  FAD $\downarrow$ & DeSync $\downarrow$ & LPAPS $\downarrow$ \\ 
         \midrule
          GT & -  & 27.4 \% & 3.2 \%  & - & - & - \\
         \midrule
         VoiceLDM& Multiple &\textbf{17.2} \% &  22.5 \% & \textbf{0.36} & 1.49 & 6.48 \\ 
         VATT & VGGSound   & 98.9 \% & 75.6 \% & 0.43 & \textbf{1.33} & 6.65\\ 
         BVS  & AS-Speech   & \underline{26.4} \%  & \textbf{21.9} \% & \underline{0.41} & \textbf{1.33} & \textbf{6.38}\\ 
         \bottomrule
    \end{tabular}
    \caption{Model comparison of video to audio with context-aware speech. VoiceLDM is a text-guided context-aware TTS model.}
    \label{tab:tts}}
\end{table}

We evaluate the effectiveness of BVS on video-to-audio with context-aware speech by incorporating a pretrained text-to-speech SSL generator in MaskGCT~\cite{wang2024maskgct}. As there are no existing methods that directly address this task, we select two most relevant baselines: VoiceLDM~\cite{lee2024voiceldm}, a text-to-context-aware TTS model that generates speech aligned with the semantic content of text prompts, and VATT~\cite{liu2024tell}, a V2A model that produces ambient sounds temporally synchronized with visual input through audio captions.
We exclude the concurrent work, DualDub~\cite{tian2025dualdub}, as it focuses on generating speech from videos featuring talking heads.
We evaluate both baseline methods with their official checkpoints, and the results are reported in Table~\ref{tab:tts}. Although VoiceLDM focuses on synthesizing context-aware speech given text prompts and is trained on both AudioSet-Speech and multiple synthesized noisy speech datasets with a CLAP text encoder~\cite{laionclap2023}, its worse DeSync and LPAPS scores suggest that it struggles to generate audios with synchronized sound effects given prompts that contain events in temporal order. Conversely, VATT is designed to produce temporally aligned audio from videos; however, the high WER and $\Delta \textnormal{WER}$ indicate VATT fails to generate intelligible speech. Compared to both baselines, BVS overcomes these limitations by generating temporally and semantically synchronized audio with intelligible speech given transcripts and videos. Notably, despite being trained solely on AudioSet-Speech, BVS achieves WER values closer to the ground truth with a lower $\Delta\textnormal{WER}$, which demonstrating the robustness of BVS in generating synchronized audio with context-aware intelligible speech. 

\begin{table}[t]
    \centering
    \renewcommand{\tabcolsep}{1.2 mm}
    {\fontsize{8pt}{9pt}\selectfont
    \begin{tabular}{c|ccccc}
         \toprule
         Method & $\Delta$WER $\downarrow$ &  FAD $\downarrow$ & DeSync $\downarrow$ & LPAPS$_{\text{S}} \downarrow$ & LPAPS$_{\text{T}}\downarrow$ \\
         \midrule
          SE + VATT &  27.8 \% & 0.39 &  \textbf{1.22} & 6.69 & 6.47\\
          BVS  &  \textbf{26.9} \% & \textbf{0.38}  & 1.32 & \textbf{6.42} & \textbf{6.28}\\
         \bottomrule
    \end{tabular}
    \caption{Experiment on immersive audio background conversion.}
    \label{tab:convert}}
    \vspace{-0.2in}
\end{table}

\subsection{Immersive Audio Background Conversion}~\label{sec:bvc}
We then demonstrate the effectiveness of BVS in immersive audio background conversion. Unlike the setting in~Sec.~\ref{sec:tts}, this task takes noisy speech as input and converts it into a new context-aware speech that is naturally integrated within the given video’s content. We randomly sample 500 video–audio pairs from the AS-filtered dataset, which serve as target video and source audio. Since no direct baselines exist, we construct a proxy baseline by first applying a Metis speech enhancement model~\cite{wang2025metis} to obtain clean speech, which is then directly mixed with sound effects generated by VATT. To prevent VATT from generating unintelligible human voices, we exclude caption inputs and make the model focus on generating sound effects. 
To measure the consistency of converted audio between source and target audio, we report LPAPS$_\text{T}$ and LPAPS$_\text{S}$, where LPAPS$_\text{T}$ measures similarity to the audio track of target video, and LPAPS$_\text{S}$ measures similarity to the source audio.
As shown in Table~\ref{tab:convert}, although the baseline achieves a lower DeSync score, our method outperforms it in terms of LPAPS and FAD, reflecting higher perceptual similarity to ground truth. In contrast, simple post hoc mixing of clean speech with generated sound effects often results in poor semantic alignment between speech and environmental context, whereas BVS produces coherent and context-aware audio.

\subsection{Ablation study}
We first validate whether the audio semantic tokens obtained from W2V-BERT have the ability to represent non-speech information. As shown in the first row of Table~\ref{tab:ablation}, we evaluate reconstruction results on AS-filtered using our VS2A model conditioned on ground-truth audio semantic tokens and videos. The metrics of reconstructed samples reflect that the VS2A model can generate audio that contains both speech and non-speech sounds, with better scores than generation results in Sec.~\ref{sec:tts} and Sec.~\ref{sec:bvc}.
We then conduct ablation studies to evaluate the importance of the speech semantic  token from the SE model and the audio semantic tokens in our two-stage method.
Firstly, we remove the audio semantic tokens from BVS, converting it into a single-stage model that directly predicts acoustic tokens, which we denote as BVS-Single. Secondly, we remove speech SSL tokens obtained from raw audios by replacing the SE speech SSL generator with a transcript-based speech SSL generator used in MaskGCT~\cite{wang2024maskgct}. We use Whisper~\cite{radford2023robust} to enable alignment of speech onset and offset times. We use filtered transcripts obtained from an ASR model by~\cite{lee2024voiceldm} to guide phonetic information. We refer to this as BVS-Text.
As illustrated in Table~\ref{tab:ablation}, both methods fail to learn speech information effectively with significantly higher WER and FAD. In BVS-Text, although the model is given transcripts with speech onset-offset timestamps, the model still fails to learn the complex alignment of speech and sound effects from video and transcripts. This indicates that the models fail to capture the relationship between video, speech, and context information. In contrast, our method with both speech SSL tokens obtained directly from raw audios, and a two-stage design explicitly guides the model to disentangle, fuse, and refine the information of speech and sound effects. 

\begin{table}[t]
    \centering
        \vspace{-0.15in}
    {\fontsize{8pt}{9pt}\selectfont
    \renewcommand{\tabcolsep}{0.8mm}
    \begin{tabular}{c|cc|cccc}
         \toprule
         Method &SP-SSL & Audio SSL & $\Delta$WER $\downarrow$ &  FAD $\downarrow$ & DeSync $\downarrow$ & LPAPS $\downarrow$ \\
         \midrule
          VS2A & -  & - & 19.8 \% & 0.25 & 1.02 & 5.12 \\ 
         \midrule
         BVS-Single &$\checkmark$ &  & 99.1 \% & 1.18 & 1.38 & 6.78\\
         BVS-Text & & $\checkmark$  & 98.7 \% & 1.12 & 1.49 & 7.01\\
         BVS (Ours) & $\checkmark$   & $\checkmark$ &  \textbf{21.9} \% & \textbf{0.41} & \textbf{1.33} & \textbf{6.38}\\
         \bottomrule
    \end{tabular}
    \caption{Ablation study. SP-SSL stands for speech semantic tokens, and audio SSL stands for unified audio semantic tokens.}
    \label{tab:ablation}}
    \vspace{-0.15in}
\end{table}

\section{Conclusion}
In this paper, we introduce BVS, which addresses the challenge of video-to-audio generation with intelligible and environmentally aware speech. By leveraging a pretrained speech SSL generator during training, BVS disentangles speech information from noisy audio, and then effectively fuses cues from video, speech, and audio modalities, and generates audio with a two-stage design. Our experiments demonstrate that BVS can be applied to diverse real-world scenarios, including both generation and conversion tasks. Despite these promising results, we observed some limitations. Since BVS relies on speech SSL models for the unified audio semantic token space, which performance on non-speech sound effects remains constrained. Moreover, due to limitations in the available datasets, accurate lip–speech synchronization remains challenging. These limitations point to important directions for future research. In particular, exploring more powerful unified speech–audio SSL representations may improve sound effect generation, while incorporating richer multimodal datasets could enhance synchronization and further advance the quality of video-to-audio synthesis.

\newpage
\bibliographystyle{IEEEbib}
\bibliography{strings}

\end{document}